\documentclass[twocolumn,prl,superscriptaddress]{revtex4-1}

\usepackage{graphicx}
\usepackage{amsmath}
\usepackage{amsfonts}
\usepackage{amssymb}
\usepackage{xcolor}
\usepackage{hyperref}
\hypersetup{colorlinks=true,allcolors=blue}

\usepackage{footnote}
\usepackage{textcomp}

\newcommand{\mbf}[1]{\mathbf{#1}}
\renewcommand{\t}[1]{\textrm{#1}}

\newcommand{\q}{\mbf{q}}

\newcommand{\g}{\gamma}

\renewcommand{\k}{\kappa}

\newcommand{\m}{\mu}

\newcommand{\w}{\omega}

\newcommand{\D}{\Delta}

\newcommand{\+}{^\dagger}
\renewcommand{\>}{\rangle}

\newcommand{\G}{\vert G\>}
\newcommand{\X}{\vert X\>}

\newcommand{\XX}{\vert X\>\<X\vert}
\newcommand{\XG}{\vert X\>\<G\vert}
\newcommand{\GX}{\vert G\>\<X\vert}

\begin{document}

\title{Emission-frequency separated high quality single-photon sources enabled by phonons}

\author{M. Cosacchi}
\affiliation{Theoretische Physik III, Universit{\"a}t Bayreuth, 95440 Bayreuth, Germany}
\author{F. Ungar}
\affiliation{Theoretische Physik III, Universit{\"a}t Bayreuth, 95440 Bayreuth, Germany}
\author{M. Cygorek}
\affiliation{Department of Physics, University of Ottawa, Ottawa, Ontario, Canada K1N 6N5}
\author{A. Vagov}
\affiliation{Theoretische Physik III, Universit{\"a}t Bayreuth, 95440 Bayreuth, Germany}
\affiliation{ITMO University, St. Petersburg, 197101, Russia}
\author{V. M. Axt}
\affiliation{Theoretische Physik III, Universit{\"a}t Bayreuth, 95440 Bayreuth, Germany}

\begin{abstract} We demonstrate theoretically that the single-photon purity of photons
emitted from a quantum dot exciton prepared by phonon-assisted off-resonant
excitation can be significantly higher in a wide range of parameters than that
obtained by resonant preparation for otherwise identical conditions. Despite the off-resonant excitation the
brightness stays on a high level.  These surprising findings exploit that the
phonon-assisted preparation is a two-step process where phonons first lead to a
relaxation between laser-dressed states while high exciton occupations are
reached only with a delay to the laser pulse maximum by adiabatically undressing
the dot states. Due to this delay, possible subsequent processes, in particular
multi-photon excitations, appear at a time when the laser pulse is almost
gone. The resulting suppression of reexcitation processes increases the single-photon purity.
Due to the spectral separation of the signal photons from the laser frequencies
this enables the emission of high quality single photons not disturbed by a
laser background while taking advantage of the robustness of the phonon assisted scheme.
\end{abstract}

\maketitle

On-demand single-photon sources continue to gain attention as
key building blocks in quantum technological applications, ranging from novel
metrology over quantum communication to  quantum computing.  Semiconductor
quantum dots (QDs) have proven to be suitable single-photon emitters
\cite{Michler2000,santori2001tri,Santori2002,He2013,Somaschi2016,wei2014det,Ding2016,Schweickert2018}
that due to their high compatibility with  existing semiconductor technology 
are promising candidates for device applications.
In contrast to atomic systems,
these nanoscale structures are prone to the influence of the surrounding solid
state crystal matrix.  Longitudinal acoustic (LA) phonons are the main source of
decoherence of excitons in semiconductor QDs even at cryogenic temperatures of a
few Kelvin \cite{Knorr2003,Vagov2007,Ramsay2010a,Kaer2010,Reiter2017}.
Nevertheless, phonon-assisted off-resonant QD excitations have been shown to provide a robust alternative
to resonant exciton preparation schemes \cite{glassl2013pro,Ardelt2014,Reiter2014,Bounouar2015,Quilter2015}.
In this letter, we demonstrate
theoretically that, quite unexpectedly, the coupling to LA phonons combined with
off-resonant driving can be extremely beneficial for a single-photon source
based on a QD-cavity system, allowing for the generation of  high-quality
single-photons that are easily detectable due to their spectral separation from the laser pulses used for the excitation of the QD.

Placing a QD in a cavity strongly enhances the photon emission by enlarging the
effective dot-cavity coupling and by setting a preferable emission axis.
When exciting the QD exciton resonantly, the frequencies of the excitation and the signal are identical - separating the two is a formidable experimental challenge.
In fact, spectral separability is achievable, e.g., by wetting layer excitation or by exciting the biexciton via the two-photon resonance and subsequently exploiting the biexciton-exciton cascade \cite{Schweickert2018,Hanschke2018}.
But while the former introduces a time jitter that reduces the on-demand character of the photon source, the latter is sensitive to small fluctuations of excitation parameters such as the laser energy and the pulse area.
Both problems are overcome by an off-resonant excitation of the quantum dot, which is thus extremely advantageous.
Indeed, it has recently been shown that the robustness of off-resonant excitation schemes paves the way to excite two spatially separated QDs with different transition energies simultaneously with the same laser pulse, which is a milestone towards the scalability of complex quantum networks \cite{Reindl2017}.

\begin{figure}[t]
	\centering
	\includegraphics[width=0.45\textwidth]{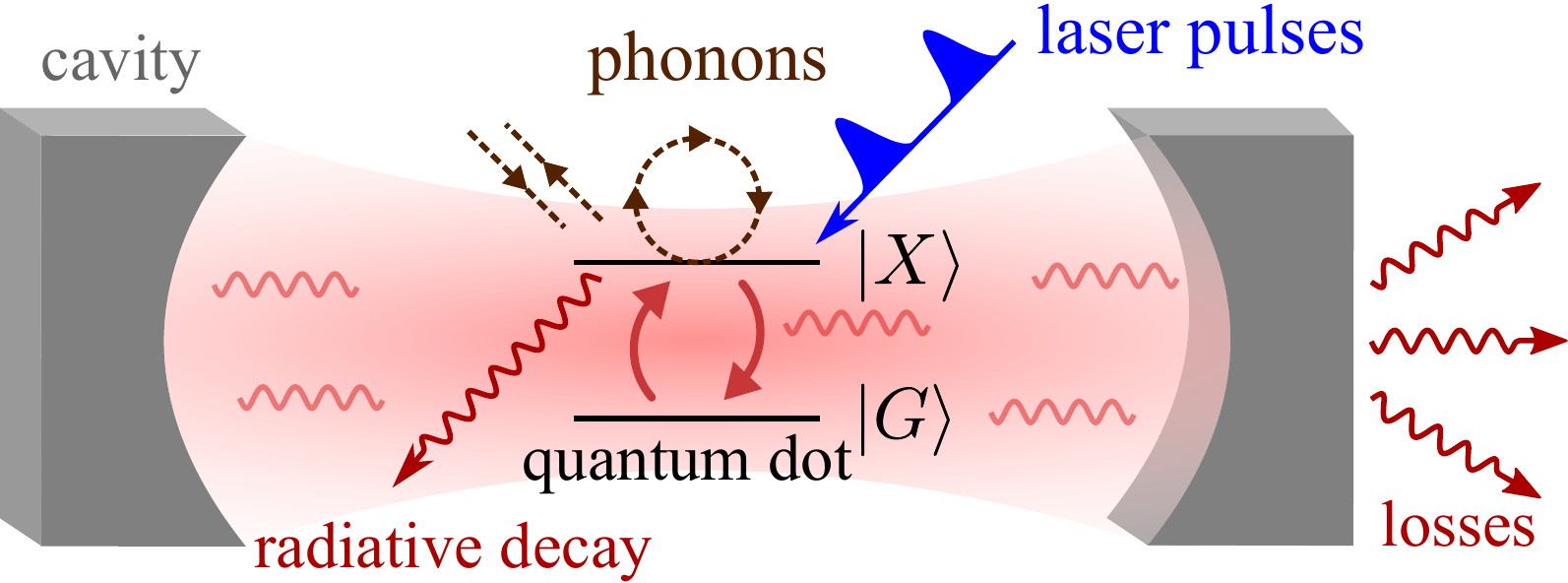}
	\caption{{Sketch of the system under consideration.}  A two-level
QD with a ground state $\G$ and an exciton state $\X$ is coupled to a lossy
single-mode microcavity. 	 The $\G\to\X$ transition is driven by external
laser pulses and the exciton state is coupled to LA phonons in a pure-dephasing
manner.  Finally, the dot can decay radiatively. }
	\label{fig:sketch}
\end{figure}

The quality of a QD-cavity system as an on-demand single-photon source is typically
quantified by several key figures of merit, such as the single-photon purity
$\mathcal{P}$ and the brightness $\mathcal{B}$.  While the former measures
whether indeed a single photon is emitted by the source, the latter
characterizes its total photon yield \cite{Somaschi2016}.  When
$\mathcal{P}=\mathcal{B}=1$, the source emits a single photon with a probability
of unity at every excitation pulse via the cavity.  The single-photon purity (SPP) can be extracted
from a Hanbury Brown-Twiss coincidence experiment
\cite{Hanbury1956,Santori2002,Bentham2016,Prtljaga2016,Weiss2016,Ding2016,Schweickert2018},
which gives a conditional probability to detect a second photon when a first one
has already been detected.  Suppressing this probability is possible, e.g., by
parametric down-conversion,  which enhances the SPP, albeit at
the cost of a severely reduced brightness  of the photon source \cite{Pan2012}.
Maximizing both SPP and brightness is of utmost importance to create
efficient single-photon emitters.  

Simultaneously large $\mathcal{P}$ and $\mathcal{B}$ in a QD-cavity system can
be achieved by exciting the dot resonantly by ultra-short laser pulses \cite{Santori2002,He2013,Ding2016}.
However, shortening the pulse duration is equivalent to widening it spectrally.
The detrimental influence of exciting higher-lying states, especially the biexciton state of the QD by short pulses is discussed in Ref.~\cite{Gustin2018}.
In view of the various advantages of phonon-assisted off-resonant excitations listed above, the question arises how photonic characteristics such as SPP and brightness perform under off-resonant schemes.
In short, we want to explore whether all of the advantages of phonon-assisted off-resonant schemes come at the cost of severely reduced photonic properties.

It is expected that driving a QD off-resonantly is much less
efficient.  For longer and stronger pulses the resulting quantum state of a
QD-cavity system contains an admixture of multi-photon states, which reduces the
SPP.  Phonon-induced dephasing is expected to degrade the
quantum state even further.  But paradoxically quite the opposite can take
place: a combination of off-resonant driving with the phonon-induced relaxation between laser-dressed QD states
leads eventually to high exciton occupations
in a subsequent adiabatic undressing process \cite{Barth2016b}.
In this letter, we demonstrate that the delay of the exciton creation caused by the undressing
suppresses  the probability for multi-photon
generation.
Therefore, comparing off-resonant and resonant excitation with otherwise same conditions may, quite unexpectedly, yield enhanced SPPs in the off-resonant case.
The best values predicted in this letter are even comparable to the best values obtained so far within resonant schemes addressing the exciton.

We model the QD-cavity system as a laser-driven two-level system with a ground
state $\G$ and an excited state $\X$, $H_{\t{DL}}=-\hbar\D\w_{\t{LX}}\XX-\frac{\hbar}{2}f(t)\left(\XG+\GX\right)$, coupled to a single-mode microcavity (cf. Fig.~\ref{fig:sketch}), $H_{\t{C}}=\hbar\D\w_{\t{CL}}a\+ a+\hbar g\left(a\+ \GX+a\XG\right)$, which is on resonance with the QD exciton.
Here, $\D\w_{\t{LX}}$ and $\D\w_{\t{CL}}$ are the laser-exciton and cavity-laser detuning, respectively, and $a$ is the photon annihilation operator in the cavity, which is coupled to the dot by the coupling constant $g$.
A train
of Gaussian pulses is assumed represented by the laser envelope function $f(t)$.
The excitation
can leave the system either via radiative decay or cavity losses modeled by Lindblad rates $\g$ and $\k$, respectively.  Finally, the
exciton is coupled to a continuum of LA phonons in a pure-dephasing manner
\cite{Krummheuer2002}, $H_{\t{Ph}}=\hbar\sum_\q \w_\q b_\q\+ b_\q+\hbar\sum_\q \left(\g_\q^{\t{X}}b_\q\+ +\g_\q^{\t{X}*}b_\q\right)\XX$.
$b_\q$ annihilates a phonon in the mode $\q$ coupled to the dot by the coupling constant $\g_\q^{\t{X}}$.
Full details of the model and of our numerical approach are
given in the supplemental material \cite{supp}\nocite{Besombes2001,Borri2001,
Vagov2011,Makri1995a,Makri1995b,McCutcheon2016}.
It is worthwhile to note that we use  path-integral methods for our simulations
that allow us to perform all simulations without approximation to the model \cite{Barth2016,Cygorek2017,Cosacchi2018,supp}.

For the calculations, standard GaAs parameters are used \cite{Krummheuer2005} for a QD of $6\,$nm diameter (for details on the phonon coupling consider the supplement \cite{supp}).
If not stated otherwise, the excitation pulse full width at half maximum is set to $7\,$ps, the cavity
mode is resonant with the QD transition, the dot-cavity coupling is $\hbar
g=50\,\m$eV,  the radiative decay rate is $\hbar\g=20\m$eV, and the cavity loss
rate is $\hbar\k=50\m$eV.  This corresponds to a Purcell factor of
$F_{\t{P}}=g^2/(\g\k)=2.5$.  The initial phonon distribution is assumed to be thermal with a temperature of $T=4.2\,$K.

The main target quantities of interest in this paper, the SPP $\mathcal{P}$ and the brightness $\mathcal{B}$,
are obtained from path-integral simulations of the two-time photonic correlation function
$G^{(2)}(t,\tau)=\<a\+(t)a\+(t+\tau)a(t+\tau)a(t)\>$ and the time dependent
photon occupation $\<a\+ a\>(t)$, respectively.
In order to express the SPP in terms of 
$G^{(2)}(t,\tau)$ one first needs to take the average over the first time argument
$t$, i.e., $G^{(2)}(\tau)=\int_{-\infty}^{\infty}dt\,G^{(2)}(t,\tau)$, which yields a
function with the delay time $\tau$ of the coincidence measurement as its single
argument.  The probability $p$ of detecting a second photon during the same
excitation pulse after a first one has already been emitted thus can be obtained
by
\begin{align}
p=\frac{\int_{-T_{\t{Pulse}}/2}^{T_{\t{Pulse}}/2}d\tau\,G^{(2)}(\tau)}{\int_{T_{\t{Pulse}}/2}^{3 T_{\t{Pulse}}/2}d\tau\,G^{(2)}(\tau)}\, ,
\end{align}
where $T_{\t{Pulse}}$ is the separation of the pulses in the pulse train.
The SPP is then defined as $\mathcal{P}=1-p$.
Note that $-\infty<\mathcal{P}\leq1$, where the lack of a lower bound is due to the possibility of bunching instead of anti-bunching.

In this work, the brightness of the source is modeled as the integrated leakage
of the average photon number during the duration of one pulse, i.e.,
$\mathcal{B}=\k\int_{-T_{\t{Pulse}}/2}^{T_{\t{Pulse}}/2}dt\, \<a\+
a\>(t)$.  Due to the definition, this quantity formally ranges in
$0\leq\mathcal{B}<\infty$ without an upper bound since in principal infinitely
many photons can exist in a single electromagnetic field mode.

\begin{figure}[t]
	\centering
	\includegraphics[width=0.45\textwidth]{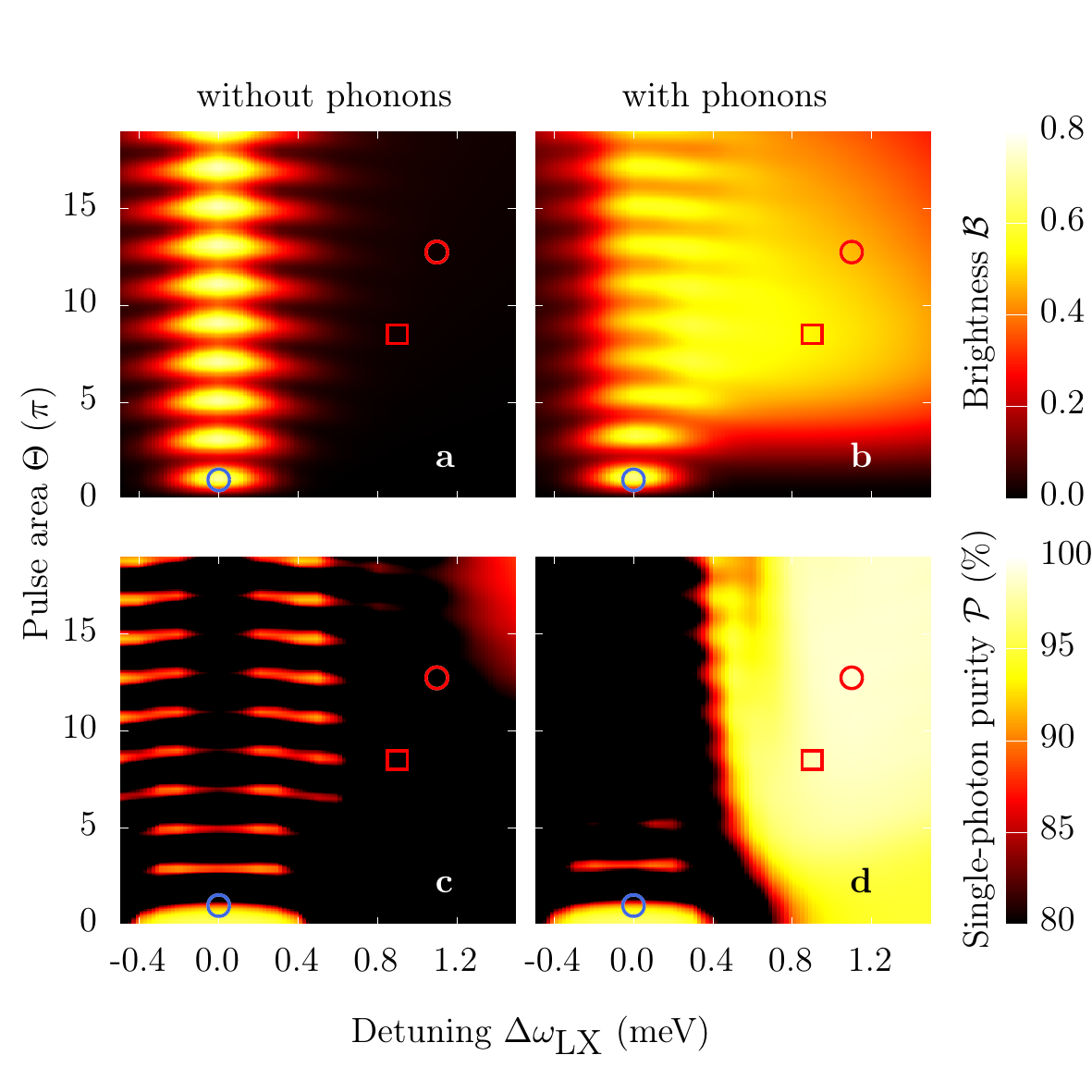}
	\caption{ Brightness $\mathcal{B}$ (panels a, b) and
SPP $\mathcal{P}$ (panels c, d) as a function of the
laser-exciton detuning $\D\w_{\t{LX}}$  and the excitation pulse area $\Theta$
of a pulse in the pulse train.  The left column (a, c) is the result of
a phonon-free calculation, the right column (b, d) includes the
coupling to a continuum of LA phonons.  Blue circle: resonant $\pi$-pulse excitation. Red
circle: maximal SPP (with phonons). Red square: optimal SPP and brightness
for off-resonant excitation (with phonons).}
	\label{fig:detareafignew}
\end{figure}

In Fig.~\ref{fig:detareafignew}a the brightness simulated without phonons is
shown as a function of the detuning $\D\w_{\t{LX}}$ between the central laser
frequency and the transition frequency connecting the ground and the exciton
state of the QD as well as the pulse area $\Theta$.  An oscillatory behavior as
a function of the pulse area with maxima at odd multiples of $\pi$ is observed
(cf. Fig.~\ref{fig:detareafignew}a).  This is a consequence of the well-known
Rabi rotation of the exciton occupation since the exciton feeds the cavity
photons, which in turn are measured by the brightness.  As a function of the
detuning, the regions of high brightness are confined to a fairly small range
around resonance.  The inclusion of phonons drastically changes this picture
(cf. Fig.~\ref{fig:detareafignew}b).  Through off-resonant excitation with
detunings that can be bridged by the emission of LA phonons, a nonvaninshing
brightness can be obtained in a previously dark region.  Note that the asymmetry
with respect to the sign of the detuning is due to the low temperature of
$T=4.2\,$K considered here where phonon absorption is largely suppressed.

\begin{figure*}[t]
	\centering
	\includegraphics[width=0.9\textwidth]{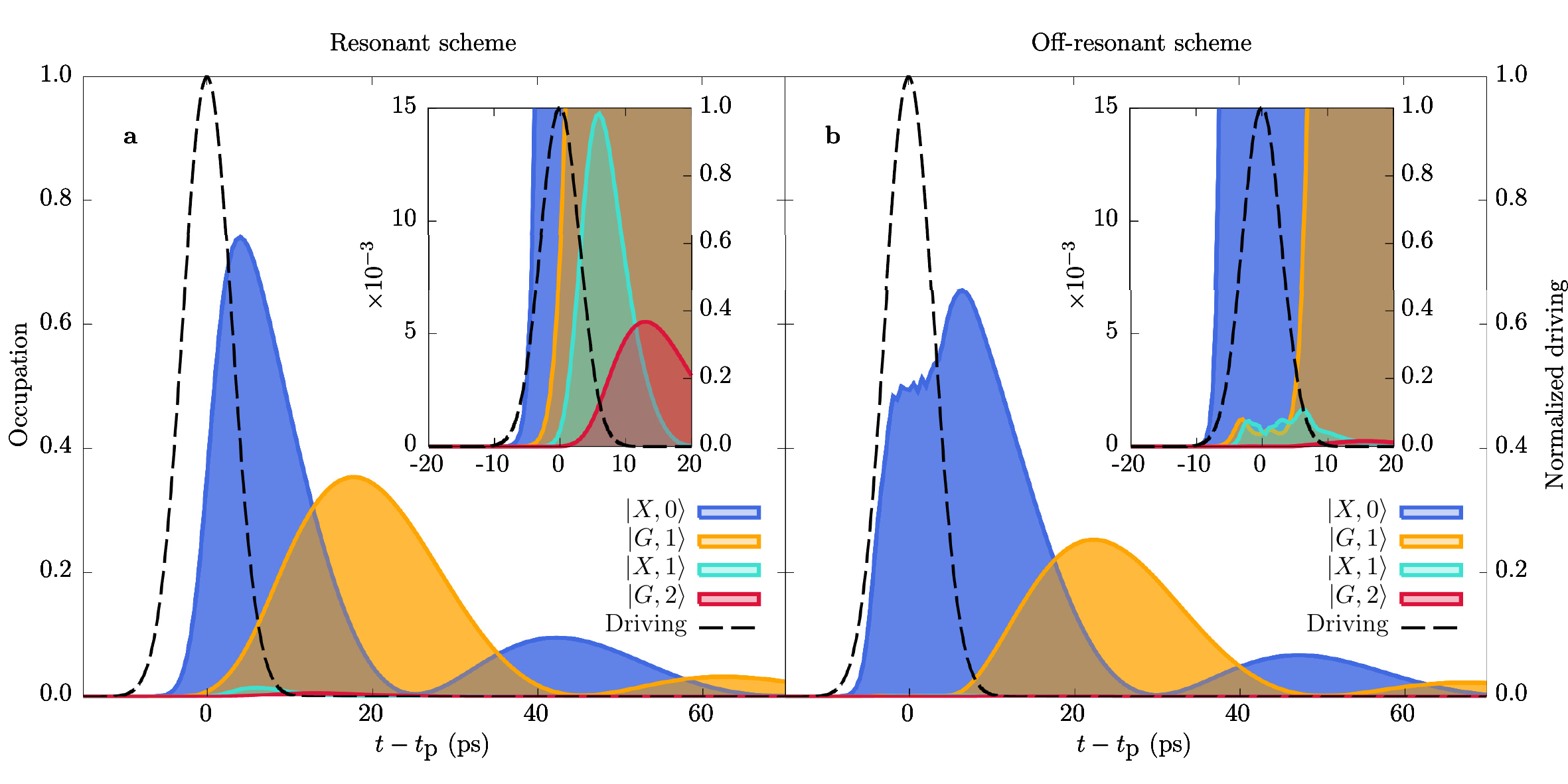}
	\caption{Time-dependent occupations:
a) after resonant $\pi$-pulse excitation (cf. blue circle in
Fig.~\ref{fig:detareafignew}) and  b)  in the
off-resonant phonon-assisted case (cf. red circle in
Fig.~\ref{fig:detareafignew}).
The occupations of the states $|X,0\>$, $|G,1\>$,
$|X,1\>$, and $|G,2\>$  are shown as colored filled curves.  The Gaussian
envelope of the laser driving pulse normalized to its maximum value centered at
$t_{\t{p}}$ is shown as a black dashed line. The insets show the same curves,
respectively, on a zoomed-in scale for the occupations.  }
	\label{fig:occupfig}
\end{figure*}

The SPP in the phonon-free case
(cf. Fig.~\ref{fig:detareafignew}c) also displays Rabi rotational behavior but
decreases with rising pulse area close to resonance, which is due to a
reexcitation of the QD during the same laser pulse.  This leads to the emission
of more than one photon per pulse, thus diminishing the SPP.
Although a SPP can always be calculated, one should be aware that it
constitutes a physically meaningful quantity only for finite brightness.
Therefore, the area of increased SPP in the upper right corner of
Fig.~\ref{fig:detareafignew}c is of no physical relevance.

It is intuitively expected that the continuum of LA phonons reduces the quantum
correlations of the system and thus the SPP by inducing a
manifold of transitions between its quantum states.  However, contrary to these
expectations Fig.~\ref{fig:detareafignew}d reveals a huge systematic increase in
$\mathcal{P}$ at $\D\w_{\t{LX}} \gtrsim 0.5\,$meV.  Moreover, the maximum
$\mathcal{P}_{\t{max}} = 98.8\%$  (red circle) is even  larger than $90.7\%$
obtained for the resonantly driven system (blue circle).  Combined with an
appreciably large $\mathcal{B}$, this indicates a possibility to have a good
quality single-photon source in the off-resonant excitation regime.  Note that
$\mathcal{B} = 0.46$ observed at the point  of $\mathcal{P}_{\t{max}} $ (cf. red
circle in  Fig.~\ref{fig:detareafignew}b) is not much smaller than the maximal
value of $0.67$ achieved in the resonantly driven case (cf. blue circle in
Fig.~\ref{fig:detareafignew}b).
It is also noteworthy that it is possible to obtain a significantly larger brightness at the cost of a slight decrease in the SSP. For example, if we choose a trade-off by maximizing the sum of the squares of the two figures of merit in the off-resonant regime, we obtain $\mathcal{B}=0.53$ and $\mathcal{P}=98.1\%$ (red square).
This value for $\mathcal{P}$ is close to typical experimental values obtained for resonant excitation of the quantum dot exciton ($98.8\%$ \cite{He2013}, $99.1\%$ \cite{Ding2016}) even though the pulse lengths in Refs. \cite{He2013,Ding2016} have been slightly shorter \footnote{Note that other system parameters in these experiments are also different from the ones used here, e.g., the Purcell factor reported in Ref.~\cite{Ding2016} is 6.3 compared with 2.5 in our case.}.


To explain the mechanism behind this observation, one needs to consider the
dynamics of the QD-cavity states.  In Fig.~\ref{fig:occupfig}, the time
dependent occupations in the resonant and the off-resonant case (cf. the blue
and red circles in Fig.~\ref{fig:detareafignew}, respectively) are compared.
The considered states are product states between the QD states and a photon
state with photon number $n$.  After resonant $\pi$-pulse excitation
(cf. Fig.~\ref{fig:occupfig}a), the exciton state $|X,0\>$ without photons is
occupied (blue curve).  The cavity coupling rotates the dot back to its ground
state and produces one photon in the cavity (orange curve).  Because the driving
is still nonzero at this point, the dot is reexcited to produce an occupation of
the state $|X,1\>$ (green curve), which is shown in the inset of
Fig.~\ref{fig:occupfig}a.  Finally, the cavity coupling leads to an occupation
of the ground state with two photons $|G,2\>$ (red curve), such that the
SPP is diminished.

In contrast to the $\pi$-pulse induced rotation of the Bloch vector, the
off-resonant excitation scheme exploits an effect called adiabatic undressing
\cite{Barth2016b}.  Switching on the laser transforms the dot states to a new
energy eigenbasis commonly known as laser-dressed states, the gap between which
can be bridged by LA phonons with typical energies of a few meV.  At low
temperatures, the lower dressed state becomes occupied via phonon emission.
However, the phonon-induced relaxation is only efficient when both dressed states have
roughly equal exciton components. Thus,  the exciton state
exhibits typically occupations of the order of 50\% after the relaxation is completed \cite{Barth2016b}.
When the laser is turned off adiabatically, the lower dressed state is
subsequently transformed to the exciton state in the original basis provided the detuning is positive
(otherwise the ground state is reached \cite{Barth2016b}).  This
adiabatic undressing of the dot states therefore boosts the exciton occupation
only at the end of the pulse (cf. the blue curve in Fig.~\ref{fig:occupfig}b).
This in turn means that during the phase of phonon relaxation no photon can be
put into the cavity efficiently (cf. the orange curve in
Fig.~\ref{fig:occupfig}b).  When finally the adiabatic undressing-induced
exciton boost occurs, the occupation of $|G,1\>$ follows
(cf. Fig.~\ref{fig:occupfig}b).  Since the excitation pulse is basically gone by
then, the reexcitation of the QD is strongly suppressed (green curve), such that
effectively no second photon can be put into the cavity (red curve).  This
implies a far higher SPP than in the resonant counterpart, as
is observed in Fig.~\ref{fig:detareafignew}d.  In summary, the delay of the
exciton occupation caused by the two-step procedure of first relaxing to a
dressed state via phonon emission and then reaching the exciton by adiabatic
undressing is responsible for the enhancement of the SPP.

\begin{figure}[t]
	\centering
	\includegraphics[width=0.5\textwidth]{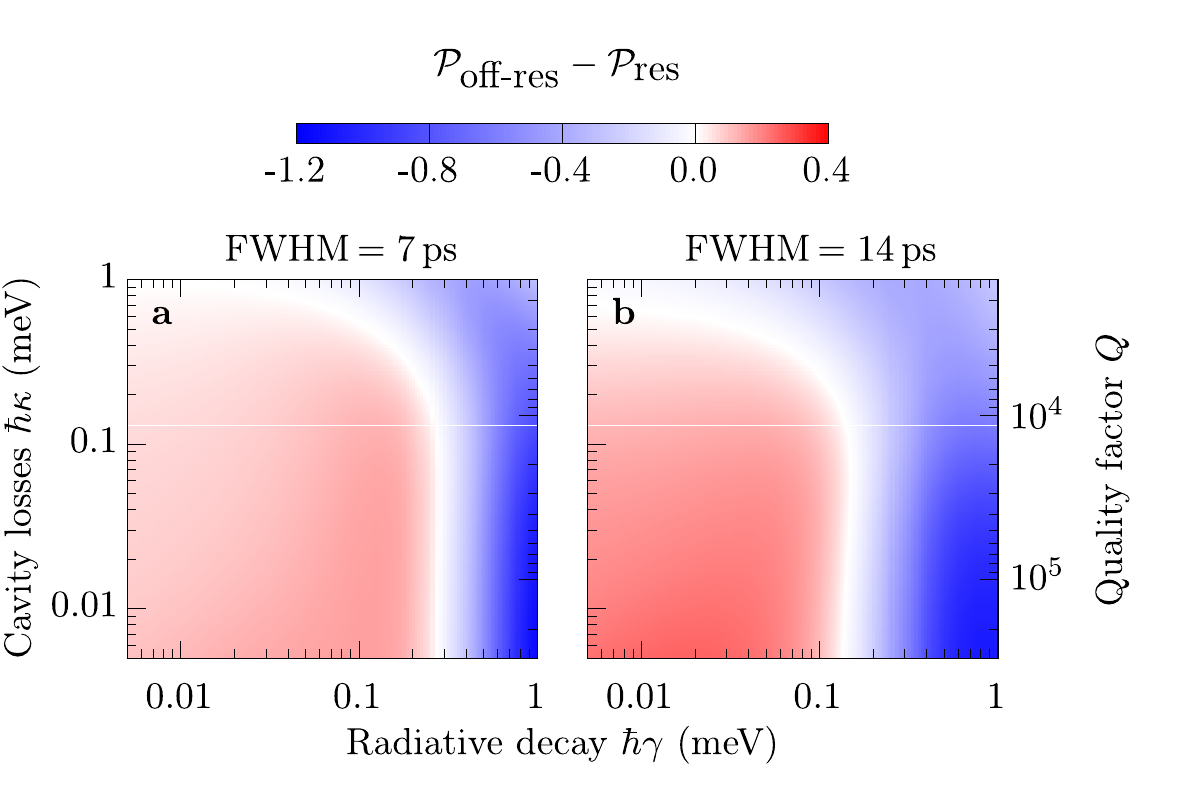}
	\caption{The difference between the SPP
after off-resonant phonon-assisted excitation $\mathcal{P}_{\t{off-res}}$ and
after resonant $\pi$-pulse rotation  $\mathcal{P}_{\t{res}}$ is shown for two
different pulse lengths (FWHM), namely: a) $7\,$ps and b)
$14\,$ps, as a function of radiative decay $\hbar\g$  and cavity losses
$\hbar\k$.  The cavity quality factor $Q=\w_{\t{c}}/\k$ is obtained via the
cavity losses assuming a cavity single-mode energy of $\hbar\w_{\t{c}}=1.5\,$eV.
The pulse area is set to $12.75\pi$ and $\D\w_{\t{LX}}=1.1\,$meV.
 }
	\label{fig:gkpurfig}
\end{figure}

To quantify the robustness of the phonon-induced SPP enhancement against
variations of other system parameters, the difference between the SPP after off-resonant excitation and after the resonant one is shown as a
function of the radiative decay $\g$ and the cavity loss rate $\k$ in
Fig.~\ref{fig:gkpurfig}.  A positive value (reddish shade) indicates a set of
parameters where the SPP is enhanced for off-resonant excitation.  We find
such an enhancement for a wide parameter regime in $\k$ and $\g$ that is
experimentally well accessible.  Also, changing the pulse length from $7\,$ps in
Fig.~\ref{fig:gkpurfig}a to $14\,$ps in Fig.~\ref{fig:gkpurfig}b does not change
the phonon-induced SPP enhancement qualitatively.  The reason why the SPP
for off-resonant excitation falls below the resonant one in the bad cavity limit
and/or in the limit of high radiative losses is that relaxation processes limit
the time available for the adiabatic undressing which eventually becomes
incomplete.

In conclusion, we have presented a seemingly paradoxical scheme for the
phonon-assisted operation of a QD-cavity system as a single-photon source, where
the excitation is spectrally separated from the generated photons.  Two factors
that would separately lead to a quality degradation - off-resonant driving and
dot-phonon coupling - in combination result in a huge boost in critical
characteristics of a single-photon source. We have demonstrated that the
achievable single-photon purity can be noticeably higher than for resonant excitation while
the brightness is still at an acceptable level. The physical mechanism of this
enhancement - the adiabatic undressing - is realized in a wide interval of
physically accessible parameters.

\acknowledgements
M.Cy. thanks the Alexander-von-Humboldt foundation for support
through a Feodor Lynen fellowship.  A.V. acknowledges the support from the
Russian Science Foundation under the Project 18-12-00429 which was used to study
dynamical processes non-local in time by the path integral approach.  This work
was also funded by the Deutsche Forschungsgemeinschaft (DFG, German Research
Foundation) - project Nr. 419036043.

\bibliography{bib}

\end{document}